# Two-Input Enzymatic Logic Gates Made Sigmoid by Modifications of the Biocatalytic Reaction Cascades

Oleksandr Zavalov,[1] Vera Bocharova,[2] Jan Halámek,[3] Lenka Halámková,[3] Sevim Korkmaz,[4] Mary A. Arugula,[5] Soujanya Chinnapareddy,[3] Evgeny Katz,[3] Vladimir Privman[1]

[1] *Department of Physics, Clarkson University, Potsdam, NY 13699*
[2] *Chemical Sciences Division, Oak Ridge National Laboratory, Oak Ridge, TN*
[3] *Department of Chemistry and Biomolecular Science, Clarkson University, Potsdam, NY*
[4] *Department of Chemical and Biomolecular Engineering, Clarkson University, Potsdam, NY*
[5] *Department of Chemistry, University of Utah, Salt Lake City, UT 84112*

**Abstract:** Computing based on biochemical processes is a newest rapidly developing field of unconventional information and signal processing. In this paper we present results of our research in the field of biochemical computing and summarize the obtained numerical and experimental data for implementations of the standard two-input **OR** and **AND** gates with double-sigmoid shape of the output signal. This form of response was obtained as a function of the two inputs in each of the realized biochemical systems. The enzymatic gate processes in the first system were activated with two chemical inputs and resulted in optically detected chromogen oxidation, which happens when either one or both of the inputs are present. In this case, the biochemical system is functioning as the **OR** gate. We demonstrate that the addition of a "filtering" biocatalytic process leads to a considerable reduction of the noise transmission factor and the resulting gate response has sigmoid shape in both inputs. The second system was developed for functioning as an **AND** gate, where the output signal was activated only by a simultaneous action of two enzymatic biomarkers. This response can be used as an indicator of liver damage, but only if both of these of the inputs are present at their elevated, pathophysiological values of concentrations. A kinetic numerical model was developed and used to estimate the range of parameters for which the experimentally realized logic gate is close to optimal. We also analyzed the system to evaluate its noise-handling properties.

## 1. Introduction

Recently, there has been considerable interest in the design and development of chemical [1-6] and biochemical [7-12] systems which implement binary logic gates and networks. These systems were investigated for novel computational processing of chemical and biochemical signals. Special attention has been paid to biomolecular systems [13,14], including those that are based on enzymatic reactions [15-18], DNA [9,19,20], or RNA [21] and even entire cells [22,23] as a new research direction in unconventional computing [24,25]. Enzymatic reactions can be used to simulate digital logic gates [26-31] and perform elementary arithmetic [32], as well as "networked" in binary logic circuits [33-36]. Processes in such systems can emulate Boolean logic and thus used as "digital" biochemical sensors with multiple inputs [37-39], e.g., in biomedical applications [29-31].

In biotechnological environments, biomolecular computing systems promise [40] design of original biosensors with multiplexing capabilities, processing several biological signals in the digital format (**0** or **1**) [39,41]. In particular, devices based on implementations of biomolecular logic have been extensively studied for possible applications in analysis of biomarkers characteristic of various pathophysiological conditions and for diagnosis of disease or injury. For example, the output chemical concentration reaching a certain logic-**1** value could signal that an "action" is needed, however the output concentration close to logic-**0** would indicate "no action" is needed. Most developed biochemical systems consider logic-**0** as zero concentrations of the signaling substances, whereas logic-**1** is defined as experimentally convenient nonzero concentrations. However, for real biomedical conditions the logic values **0** and **1** of concentrations (or possibly ranges of concentrations) should correspond to normal physiological and pathophysiological states of the organism, respectively. In most cases the difference between the inputs at the logic **0** and **1** is comparable with the level of natural noise, making the discrimination of the **0** and **1** output signals difficult unless careful optimization of the signal processing systems is performed [42].

Noise amplification tendency of the simplest biochemical processes has to be taken into consideration in designing of biochemical process as "devices" and networks. There are several sources of noise in biochemical systems, in addition to the noise in the actual input data. However, the noise amplification is controlled by the transmission of random fluctuations of the input concentrations to the output response [43]. The amplification of the noise can build up from one step to another, preventing connection of these systems in a network for information processing [44]. Thus, suppression of the noise amplification from the input(s) to output is crucial.

A typical response in biochemical systems has a convex shape with an initial linear region, followed by saturation caused by the limitation of reaction rates and the amount of input catalysts and reagents. Generally, in order to avoid the build-up of noise, it is useful to pass the signals through biochemical filters, which give the sigmoid response profile with zero or small slopes close to the "logic point" values. Such response is quite difficult to achieve in standard biochemical reactions catalyzed by enzymes [45-47], despite the fact that the response of sigmoid shape is often observed in nature [48-50]. Figure 1 (the figures are appended after the Nomenclature summary, as Pages 21-25) illustrates the response function of an **OR** gate, as an example of a system with high noise amplification levels. This system has its maximum noise transmission factor about 690% at the logic-**00** point, i.e., sevenfold noise amplification. This transmission factor can be estimated for smooth gate-response surfaces by calculating the absolute value of the gradient in terms of the logic variables, to be defined below. Therefore, despite the fact that the system response function visually may look like a reasonable **OR** gate (see Figure 1), without an additional filtering step the system is not appropriate for network applications because of such a large noise amplification.

Novel biochemical reactions with an added "filter" have recently been designed [45,51-54,57,58] and optimized as separate elements for inclusion as "logic gates" in biochemical logic networks. Integration of a filtering step into a digital gate process allows to significantly improve the parameters and performance of the latter. This applies particularly to the behavior near the binary values **0** and **1** in the output response, corresponding, for instance, to the normal physiological and pathophysiological concentration of biomarkers [51] for liver dysfunction, abdominal trauma, and soft tissue injury.

Quite often it is necessary to analyze the information simultaneously from two biomarkers [55,56]. In such cases, logic gates can be applied, such as **OR**, **AND**, etc., to facilitate making the appropriate decision, depending on the presence or absence of each biomarker or various biomarker combinations. In recent papers [57,58], we have demonstrated implementations and analyses of such biochemical systems based on enzymatic reactions with two inputs. The common feature of these works has been the incorporation of the filtering processes, which have enabled to achieve a double-sigmoid shape of the output response (sigmoid in both inputs). It is important to develop and test various biochemical filtering approaches within a "toolbox" of binary gates, including **OR** and **AND**.

In this paper we summarize the investigation of the first reported realizations of an **AND** gate and an **OR** gate in enzymatic filtered systems with two inputs. When discussing the **AND** gate system, we will focus on the most studied case in this context [51], that of a pair of liver-injury biomarkers [55,56], the enzymes alanine transaminase (ALT) and lactate dehydrogenase (LDH). Elevated levels of both enzymes simultaneously have been used as an indication of liver injury [59,60]. We consider a model system [31] in aqueous solution rather than in serum [61]. The aqueous solution system has not only been realized (as the **AND** gate) to optically detect the presence of both enzymes, but has also been coupled to a signal-responsive polymer-brush thin-film deposited on an electrode [62]. The practical implementation of such a system requires large gate times, for which the gate realization precision by the original enzymatic cascade was low. The added filtering process not only reduced noise transmission, but also—for the larger gate times—improved the precision of the **AND** binary logic realization [51].

In the analysis and consideration of the **OR** gate implementation, we describe biocatalytic reactions which produce changes in pH [63-67], coupled with a pH-buffering mechanism [53]. The latter is performed by the addition of an appropriate amount of buffer for the "filtering" effect. Experimental results [58] demonstrated that the addition of the appropriately designed filtering step changes the gate response to the double-sigmoid shape.

We have developed and applied a kinetic modeling approach specially designed for the study of the binary-logic gate-response system properties. This numerical model, applied to the experimental data, was used for a quantitative evaluation of the sources of noise, input-to-output noise transmission factors [43,52], and tolerance properties of the realized **AND** [57] and **OR** gates [58] as components for biochemical logic and networks. Section 2 outlines the experimental procedure of an **OR** gate implementation with the double-sigmoid response. In Section 3, we present a recent realization of a biochemical **AND** gate logic function with the double-sigmoid response shape. In Section 4, we detail the kinetic modeling of logic gate functions. Finally, in Section 5 we present results of the experimental data fitting using the numerical model, and discuss the gate optimization and noise reduction approach.

## 2. OR gate function realization with double-sigmoid response

In this section we describe the **OR** gate realization designed in a biochemical system outlined in Scheme 1 (the schemes are appended at the end of this preprint as Pages 26-27). This system includes two enzymatic reactions, and also the buffering process. The enzyme biocatalysing the first process is esterase, of concentration to be denoted $E(t)$, where $t$ is the process time. This first enzymatic part has two inputs: Input 1 is ethyl butyrate (concentration $A(t)$), Input 2 is methyl butyrate (concentration $B(t)$). Here esterase reacts with either one of them, biocatalyzing the production of ethanol and methanol, respectively, and also of butyric acid, $U(t)$, as a byproduct of these processes. The increase in the butyric acid concentration in the solution leads to a decrease of pH from its initial value of $pH(t = 0) = 9.0$. The pH of the system lowers to the value which depends on the initial concentrations of the inputs, $A(0)$ and $B(0)$, of Tris buffer (concentration $T(0)$), and on other process parameters, but no less than to 4.2. The Tris buffer, $T(t)$, which is added at the initial moment of time, as $T(0)$ moles per unit volume, is the main component for the filtering process, as will be described below.

The experiments were carried out in several stages, first without buffering ($T(0) = 0$ mM), see Figure 1. As expected, the output values are equal to **0** at the logic **00**, and are close to each other at the logic points **01**, **10**, **11**, defining the binary **1** for the output. Next, we present a buffered system yielding a high-quality **OR** gate, with $T(0) = 4$ mM, Figure 2, and finally a system with excessive buffering which deteriorates the **OR** gate realization: $T(0) = 8$ mM,

Figure 3. Adding the Tris buffer, T(0), into the solution instantly leads to the production of its protonated form $TH^+$. The process with the Tris buffer involves the absorption of hydrogen ions from the solution and, as a consequence, prevention of the decrease in pH. This mild alkaline buffering effect persists only in the range of pH from 9 to value about 7.2, and only as long as the Tris buffer is not completely turned into the protonated form. Therefore, controlling the initial buffer concentration, T(0), we can delay the effect of the pH decrease, which is due to the process of continuous production of butyric acid with $(pK_a)_{\text{butyric acid}} = 4.82$ [68].

At first, when the pH is high, the second enzyme laccase (of concentration $L(t)$, see Scheme 1) is practically not active. However, as the pH of the system decreases past approximately 8.1, biocatalytic activity of laccase increases. Using $K_4Fe(CN)_6$ as a co-substrate, this enzyme then produces $K_3Fe(CN)_6$ with increasing rate. The concentrations of $K_4Fe(CN)_6$ and $K_3Fe(CN)_6$ will be denoted $F(t)$ and $P(t)$, respectively. The output, which is the product of the entire process, $P(t_g)$, is measured at the gate time $t_g$ = 800 s by absorbance (denoted Abs) at $\lambda = 420$ nm. For the present system, $Abs(t)$ and $P(t)$ are numerically almost identical (but have different units) [58]. For our modeling of the biochemical logic-gate system, here we take the physical zeros, $A_0(0) = B_0(0) = 0$, as the signals' binary **0**s, and experimentally convenient input values $A_1(0) = 10$ mM and $B_1(0) = 10$ mM as logic **1**s. Logic **1** of the output, $P_1(t_g)$, is set by the gate function itself.

For the considered **OR** gate system implementation of [58] to be accurate, the appropriate output values close to **0**, **1**, **1**, **1**, at the four logic-input combinations, **00**, **01**, **10**, **11**, respectively, should be realized. However, for the study of the noise control, the behavior of the system response should be investigated [16,17,43] not only at, but also around the logic input values. The latter study is carried out in terms of the scaled, binary-range (zero to one) dimensionless variables,

$$x=A(0)/A_1(0); \quad y=B(0)/B_1(0); \quad z(x,y)=P(t_g)/P_1(t_g), \qquad (1)$$

Note that in practical applications the logic one values of the inputs will be determined [43] by the environment in which the gate is used, but the logic zero values of the inputs (and output) need not be exactly at the physical zeros [42,43,57].

### 3. AND logic gate with double-sigmoid response

Scheme 2 shows the sequence of biocatalytic processes involved in the **AND** gate function, catalyzed by the two input enzymes ALT and LDH, and the added filter process catalyzed by an additional enzyme, glucose-6-phosphate dehydrogenase (G6PDH). Additional details of the system functioning [51,57] are presented below. We study a biocatalytic cascade to detect the high levels of the two enzyme inputs, ALT and LDH. Their simultaneous increase in concentrations, from normal to pathophysiological levels, provides [59,60] an evidence of liver injury. The sequence of the biochemical processes is shown in the "gate" section of Scheme 2:

$$\alpha\text{-KTG} + \text{Ala} \rightarrow \text{Glu} + \text{Pyr} \qquad (2)$$

$$\text{Pyr} + \text{NADH} \rightarrow \text{Lac} + \text{NAD}^+ \qquad (3)$$

Oxidation of NADH is followed by measuring the change in the absorbance, yielding the concentration of the output, β-nicotinamide adenine dinucleotide (NAD$^+$), which is produced only in the presence of the two input enzymes (and the "gate machinery" reactants α-KTG, Ala, NADH). Note that without the added filter process, we would obtain a response function that would not have a sigmoid shape in any direction. Here Glu abbreviates glutamate, Pyr stands for pyruvate, and Lac for lactate. Figure 4 illustrates the result for the "gate time" of $t_g = 600$ s, relevant for applications involving signal-responsive membranes [57,62].

As mentioned above, adding the filter will cause a delay in the signal increase and the appearance of the sigmoid shape. The "filter" section of Scheme 2 shows the added filtering process. Our filtering component of the **AND** logic gate function is based on enzyme glucose-6-phosphate dehydrogenase (G6PDH). The biocatalytic reaction is shown in the following:

$$\text{Glc6P} + \text{NAD}^+ \rightarrow \text{NADH} + \text{6-PGluc} \qquad (4)$$

This reaction has two inputs: glucose-6-phosphate (Glc6P) and cofactor ($NAD^+$), and one output: the reduced cofactor (NADH); the other reaction product in aqueous solution is 6-phosphogluconic acid (6-PGluc). The system now displays double-sigmoid behavior in both inputs ALT and LDH (Figure 5).

The output signal, $[NAD^+]$, was quantified by detecting the decrease in the concentration of NADH, measured optically at $\lambda = 340$ nm. It was calculated using the extinction coefficient, 6.22 mM$^{-1}$cm$^{-1}$, for NADH [69]. In order to study the 3D response surface of the **AND** gate function (Figures 4, 5), it is necessary to obtain the output signal values, $[NAD^+](t_g)$, not only for the inputs at the logical points **0** and **1**, which were $[ALT]_0(0) = 0.02$ U/mL, $[ALT]_1(0) = 2$ U/mL for Input 1 and $[LDH]_0(0) = 0.15$ U/mL, $[LDH]_1(0) = 1$ U/mL for Input 2, but also for varying input concentrations. Here and below, the subscripts **0** and **1** refer to the logic-point values. However, the actual noise-property analysis is best carried out in terms of the rescaled variables, defined as

$$x = \big([ALT](0)\text{-}[ALT]_0(0)\big)/\big([ALT]_1(0)\text{-}[ALT]_0(0)\big) \tag{5}$$

$$y = \big([LDH](0)\text{-}[LDH]_0(0)\big)/\big([LDH]_1(0)\text{-}[LDH]_0(0)\big) \tag{6}$$

$$z = \big([NAD^+](t_g)\text{-}[NAD^+]_0(t_g)\big)/\big([NAD^+]_1(t_g)\text{-}[NAD^+]_0(t_g)\big) \tag{7}$$

These equations (5-7) determine the normalized variables for the first input ($x$), second input ($y$), and output ($z$). Experimentally, we can map out the gate response surface, $z(x, y)$.

As shown in our recent papers [43,44,57,58], the design of a large network, which includes several biochemical logic systems, requires careful control of the built-in noise level. The noise level depends generally on the gate's environment. In order to ensure stable operation of complex networks with scalable processes, the degree of the noise amplification should be kept under control. It is desirable to have a response function, $z(x, y)$, which has small absolute values of gradients at all the logic points. The noise scaling properties of the logic gates can be estimated from the absolute values of gradients, $|\vec{\nabla}z(x,y)|_{00,01,10,11}$, of the function $z(x,y)$ at the logic points in the (*x*,*y*) plane. If the absolute value of the gradient is less than one, then the spread of noise in the logic gate system is suppressed. However, as mentioned above, without the inclusion of a properly designed filtering process it is difficult to obtain such a response function. Typically, without filtering the response function always has a gradient greater than 1 (see Figure 4), and, consequently, amplifies the noise level. Our results [57] with filtering have yielded a realization of an **AND** gate function with double-sigmoid filter response in two inputs; see Figure 5.

### 4. Kinetic modeling of the logic gate functions

In this section we outline how kinetic modeling can be applied to gate optimization. As mentioned above, the addition of the filtering process in the biocatalytic system allows an elimination of the built-in noise amplification in the response function. Therefore the logic gate can be used for designing large biochemical networks. We focus on the study of the properties and the analysis of such processes by a kinetic modeling approach. Consider as an example the **OR** gate function introduced earlier. The biocatalytic function of esterase can be described using the standard schemes

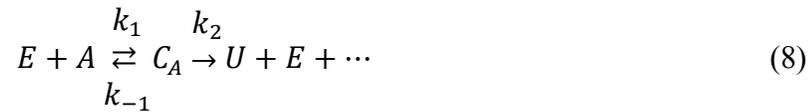

$$E + A \underset{k_{-1}}{\overset{k_1}{\rightleftarrows}} C_A \overset{k_2}{\rightarrow} U + E + \cdots \qquad (8)$$

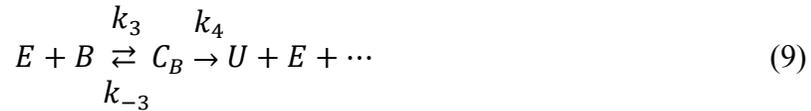

$$E + B \underset{k_{-3}}{\overset{k_3}{\rightleftarrows}} C_B \overset{k_4}{\rightarrow} U + E + \cdots \qquad (9)$$

where $C_A$ and $C_B$ are (time-dependent) concentrations of complexes. As the pH of the system changes over a wide range, the effective rate constants in eq. (8) and (9) will also change. As in the earlier work [53], we described this pH dependence by assuming fast acidification
– 9 –

equilibrium, e.g., $k_1 = \bar{k}_1 K_E/([H^+](t) + K_E)$, with $\bar{k}_1 = 1.018$ mM$^{-1}$sec$^{-1}$ [70]. Similar expressions were used for the other rate constants [70], with the values $\bar{k}_2 = 1.603$ sec$^{-1}$, $\bar{k}_3 = 0.639$ mM$^{-1}$sec$^{-1}$, and $\bar{k}_4 = 3.990$ sec$^{-1}$. The acidification equilibrium constant,

$$K_E = [E][H^+]/[EH^+] = 0.0068 \text{ mM} \tag{10}$$

and also the values for $K_{C_A} = 0.026$ mM and $K_{C_B} = 0.039$ mM were taken from the literature [53,70,71]. Thus, the effective rate constants become time-dependent via pH($t$). Next, we set up rate equations for the chemicals involved in the "esterase" part of the cascade. Here we only show one of these equations for illustration,

$$\frac{dC_B(t)}{dt} = +k_3 B(t)E(t) - k_4 C_B(t) \tag{11}$$

The buffering step of the kinetics is actually practically instantaneous. The conventional estimates for buffer functioning cannot be used because of the large range of pH variation (from 9 to about 4.2). The charge-balance equations, as well as dissociation equilibrium equations for each of the relevant chemicals, were therefore included in the overall set of equations solved numerically. Specifically, for Tris buffer [72] we have $(pK_a)_{\text{Tris}} = 8.06$. The processes biocatalyzed by laccase can be schematically described as follows [73]:

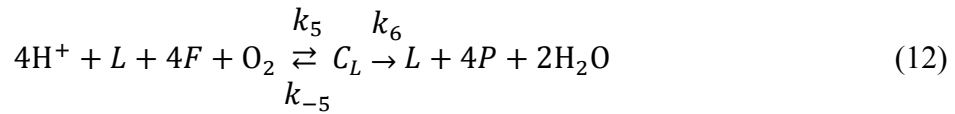

$$4H^+ + L + 4F + O_2 \underset{k_{-5}}{\overset{k_5}{\rightleftarrows}} C_L \overset{k_6}{\rightarrow} L + 4P + 2H_2O \tag{12}$$

The actual mechanism of action of laccase is more complicated than that schematically shown in eq. (12): A more detailed description of the kinetic model is explained in our earlier paper [58]. We can now write the set of rate equations for the concentrations of the various chemicals involved, for example:

$$\frac{dL}{dt} = -\tilde{k}_5(\text{pH}(t))L(t)F(t) + \tilde{k}_6(\text{pH}(t))C_L(t) \tag{13}$$

A similar approach was used for the simulation of the biochemical **AND** gate logic system. A set of kinetic differential equations was based on chemical reactions described by the eq. (2-4), and was numerically solved to study the kinetic processes in the system.

5.  **Results and Discussion**

The kinetic numerical model can be used for investigations of the biochemical system not only with the filter (**OR** gate, Figures 2, 3 and **AND** gate, Figure 5), but also without the filter (with the same rate parameter values: **OR** gate, Figure 1 and **AND** gate, Figure 4). In the figures, the experimental data are shown as compared to a model calculation. We focus on the results at the selected gate time. Using this model we have received a good-quality semi-quantitative fit for the measured experimental data, with the exception, perhaps, of those cases when the concentrations of inputs are close to the points of logical **0**, which is explained generally by the high level of relative noise for small concentrations. As described in earlier works [42-46,57,58], estimation of noise-handling properties and quality of the realized biochemical systems as "logic gates" for information processing needs a description of the system response function $z(x,y)$ using the logic variable ranges. The positions of the logical points are usually chosen on the basis of the relevant physiological conditions. For example, the non-filtered **AND** function (Figure 4) has insufficient quality to determine with satisfactory accuracy the position of a logical zero of the output, $[\text{NAD}^+]_0(t_g)$, as the value at the **00** input. Therefore, for the parameters used for our "filtered" experiment, see Figure 5, we used the calculated values [57] from our numerical model, $[\text{NAD}^+]_0(t_g) \simeq 0.2 \cdot 10^{-2}$ mM and $[\text{NAD}^+]_1(t_g) \simeq 29.8 \cdot 10^{-2}$ mM. The actual definition of the expected values or ranges of the system outputs corresponding to logic **1** may vary according to the logic gate application. As mentioned earlier, the systems without the filter (Figures 1, 4) are characterized by an appreciable increase of the input signal noise upon processing by the gate to yield the output.

The correctly filtered **OR** gate (Figure 2) and **AND** gate (Figure 5) systems do not have this problem, because the added filtering process makes the noise transmission factor at all four logic points into actual noise suppression. For example, the largest transmission factor is approximately 14% for the T(0) = 4 mM buffer concentration (**OR** gate realization, Figure 2) and it is about 8% for the realized filtering **AND** gate system (Figure 5), which is safely below 100%. The third filtered **OR** gate system (Figure 3), T(0) = 8 mM, has minor noise amplification, about 113%, and the gate realization is inaccurate: this system is an example of undesirable "overfiltering."

As expected for a good-quality filter of **AND** gate and **OR** gate logic systems, at the logic-point values the slopes $|\vec{\nabla}z(x,y)|$, in terms of the rescaled "logic-range" variables, are well below 1 (below 100%), resulting in noise suppression. Note that we can estimate the quality of the filter using the "noise tolerance" parameter specifying the extent of the possible input noise near the logical points. For the non-filtered system, the noise tolerance is zero, because it is clear from Figure 1 and Figure 4 that the largest slopes of these gate-response functions are significantly larger than 1. This is a common property of most biocatalytic reactions [15,43,52]. The non-filtered **AND** and **OR** gates will significantly amplify any input noise. Numerical estimates have shown the value of $|\vec{\nabla}z(x,y)| \approx 4.5$ (indicating input-to-output noise amplification by a factor of approximately 450%) for the **AND** gate and 6.94 for the **OR** gate, respectively. The main reason to adding the filter processes has been to reduce the slopes at all the logic points to well below 1 and to maximize the noise tolerance.

To conclude, we reviewed the experimental realization and numerical performance analysis of enzymatic **OR** [58] and **AND** [57] gate systems with a noise-reducing double-sigmoid response in two of the inputs. For the investigated enzymatic reactions, we identified the regimes of the experimental conditions necessary to have a relatively small degree of analog noise amplification.

The authors gratefully acknowledge funding of this research by the U. S. National Science Foundation (via awards CCF-1015983 and CBET-1066397 at Clarkson University), and sponsorship by the Laboratory Directed Research and Development Program of Oak Ridge National Laboratory (managed by UT-Battelle, LLC, for the U. S. Department of Energy).

**References**

[1] Credi, A. (2007). Molecules that make decisions. *Angew. Chem. Int. Ed.* **46**, 5472-5475.

[2] De Silva, A.P., Uchiyama, S., Vance, T.P., Wannalerse, B. (2007). A supramolecular chemistry basis for molecular logic and computation. *Coord. Chem. Rev.* **251**, 1623-1632.

[3] Pischel, U. (2007). Chemical approaches to molecular logic elements for addition and subtraction. *Angew. Chem. Int. Ed.* **46**, 4026-4040.

[4] Szacilowski, K. (2008). Digital information processing in molecular systems. *Chem. Rev.* **108**, 3481-3548.

[5] Andreasson, J., Pischel, U. (2010). Smart molecules at work-mimicking advanced logic operations. *Chem. Soc. Rev.* **39**, 174-188.

[6] Pischel, U. (2010). Digital operations with molecules - Advances, challenges, and perspectives. *Austral. J. Chem.* **63**, 148-164.

[7] Ashkenasy, G., Ghadiri, M.R. (2004). Boolean logic functions of a synthetic peptide network. *J. Am. Chem. Soc.* **126**, 11140-11141.

[8] Benenson, Y., Gil, B., Ben-Dor, U., Adar, R., Shapiro, E. (2004). An autonomous molecular computer for logical control of gene expression. *Nature* **429**, 423-429.

[9] Stojanovic, M.N., Stefanovic, D., LaBean, T., Yan, H. (2005). Computing with nucleic acids. In: Willner, I., Katz, E. (Eds.), *Bioelectronics: From Theory to Applications*, Wiley-VCH, Weinheim, pp. 427-455.

[10] Shapiro, E., Gil, B. (2007). Biotechnology - Logic goes in vitro. *Nature Nanotechnol.* **2**, 84-85.

[11] Win, M.N., Smolke, C.D. (2008). Higher-order cellular information processing with synthetic RNA devices. *Science* **322**, 456-460.

[12] Benenson, Y. (2009). Biocomputers: from test tubes to live cells. *Molecular Biosystems* **5**, 675-685.

[13] Benenson, Y. (2012). Biomolecular computing systems: principles, progress and potential. *Nature Rev. Genetics* **13**, 455-468

[14] Adar, R., Benenson, Y., Linshiz, G., Rosner, A., Tishby, N., Shapiro, E. (2004). Stochastic computing with biomolecular automata. *Proc. Natl. Acad. USA* **101**, 9960-

9965.

[15] Katz, E., Privman, V. (2010). Enzyme-based logic systems for information processing. *Chem. Soc. Rev.* **39**, 1835-1857.

[16] Katz, E. (2012). Molecular and Supramolecular Information Processing: From Molecular Switches to Logic Systems. (Ed.), Willey-VCH, Weinheim, (ISBN-10: 3-527-33195-6).

[17] Katz, E. (2012). Biomolecular Computing: From Logic Systems to Smart Sensors and Actuators. (Ed.), Willey-VCH, Weinheim, (ISBN-10: 3-527-33228-6).

[18] Unger, R., Moult, J. (2006). Towards computing with proteins. *Proteins* **63**, 53-64.

[19] Ezziane, Z. (2006). DNA computing: applications and challenges. *Nanotechnology* **17**, R27-R39.

[20] Margulies, D., Hamilton, A.D. (2009). Digital analysis of protein properties by an ensemble of DNA quadruplexes. *J. Am. Chem. Soc.* **131**, 9142-9143.

[21] Rinaudo, K., Bleris, L., Maddamsetti, R., Subramanian, S., Weiss, R., Benenson, Y. (2007). A universal RNAi-based logic evaluator that operates in mammalian cells. *Nat. Biotechnol.* **25**, 795-801.

[22] Tamsir, A., Tabor, J.J., Voigt, C.A. (2011). Robust multicellular computing using genetically encoded NOR gates and chemical 'wires'. *Nature* **469**, 212-215.

[23] Li, Z., Rosenbaum, M.A., Venkataraman, A., Tam, T.K., Katz, E., Angenent, L.T. (2011). Bacteria-based AND logic gate: A decision-making and self-powered biosensor. *Chem. Commun.* **47**, 3060-3062.

[24] Calude, C.S., Costa, J.F., Dershowitz, N., Freire, E., Rozenberg, G. (2009). Unconventional Computation. Lecture Notes in Computer Science, (Eds.), Vol. 5715, Springer, Berlin.

[25] Adamatzky, A., De Lacy Costello, B., Bull, L., Stepney, S., Teuscher, C. (2007). Unconventional Computing, (Eds.), Luniver Press, UK.

[26] Baron, R., Lioubashevski, O., Katz, E., Niazov, T., Willner, I. (2006). Logic gates and elementary computing by enzymes. *J. Phys. Chem. A* **110**, 8548-8553.

[27] Strack, G., Pita, M., Ornatska, M., Katz, E. (2008). Boolean logic gates using enzymes as input signals. *ChemBioChem* **9**, 1260-1266.

[28] Zhou, J., Arugula, M.A., Halámek, J., Pita, M., Katz, E. (2009). Enzyme-based universal NAND and NOR logic gates with modular design. *J. Phys. Chem. B* **113**, 16065-16070.

[29]  Manesh, K.M., Halámek, J., Pita, M., Zhou, J., Tam, T.K., Santhosh, P., Chuang, M.-C., Windmiller, J.R., Abidin, D., Katz, E., Wang, J. (2009). Enzyme logic gates for the digital analysis of physiological level upon injury. *Biosens. Bioelectron.* **24**, 3569-3574.

[30]  Pita, M., Zhou, J., Manesh, K.M., Halámek, J., Katz, E., Wang, J. (2009). Enzyme logic gates for assessing physiological conditions during an injury: Towards digital sensors and actuators. *Sens. Actuat. B* **139**, 631-636.

[31]  Halámek, J., Windmiller, J.R., Zhou, J., Chuang, M.-C., Santhosh, P., Strack, G., Arugula, M.A., Chinnapareddy, S., Bocharova, V., Wang, J., Katz, E. (2010). Multiplexing of injury codes for the parallel operation of enzyme logic gates. *Analyst* **135**, 2249-2259.

[32]  Baron, R., Lioubashevski, O., Katz, E., Niazov, T., Willner, I. (2006). Elementary arithmetic operations by enzymes: A model for metabolic pathway based computing. *Angew. Chem. Int. Ed.* **45**, 1572-1576.

[33]  Niazov, T., Baron, R., Katz, E., Lioubashevski, O., Willner, I. (2006). Concatenated logic gates using four coupled biocatalysts operating in series. *Proc. Natl. Acad. Sci. USA* **103**, 17160-17163.

[34]  Strack, G., Ornatska, M., Pita, M., Katz, E. (2008). Biocomputing security system: Concatenated enzyme-based logic gates operating as a biomolecular keypad lock. *J. Am. Chem. Soc.* **130**, 4234-4235.

[35]  Privman, M., Tam, T.K., Pita, M., Katz, E. (2009). Switchable electrode controlled by enzyme logic network system: Approaching physiologically regulated bioelectronics, *J. Am. Chem. Soc.* **131**, 1314-1321.

[36]  Tam, T.K., Pita, M., Katz, E. (2009). Enzyme logic network analyzing combinations of biochemical inputs and producing fluorescent output signals: Towards multi-signal digital biosensors. *Sens. Actuat. B* **140**, 1-4.

[37]  Windmiller, J.R., Strack, G., Chuan, M.-C., Halámek, J., Santhosh, P., Bocharova, V., Zhou, J., Katz, E., Wang, J. (2010). Boolean-format biocatalytic processing of enzyme biomarkers for the diagnosis of soft tissue injury. *Sens. Actuat. B* **150**, 285-290.

[38]  May, E.E., Dolan, P.L., Crozier, P.S., Brozik, S., Manginell, M. (2008). Towards de novo design of deoxyribozyme biosensors for GMO detection. *IEEE Sens. J.* **8**, 1011-1019.

[39]  Wang, J., Katz, E. (2010). Digital biosensors with built-in logic for biomedical

applications - biosensors based on biocomputing concept. *Anal. Bioanal. Chem.* **398**, 1591-1603.

[40] Kahan, M., Gil, B., Adar, R., Shapiro, E. (2008). Towards molecular computers that operate in a biological environment. *Physica D* **237**, 1165-1172.

[41] Wang, J., Katz, E. (2011). Digital biosensors with built-in logic for biomedical applications. *Isr. J. Chem.* **51**, 141-150.

[42] Melnikov, D., Strack, G., Zhou, J., Windmiller, J.R., Halámek, J., Bocharova, V., Chuang, M.-C., Santhosh, P., Privman, V., Wang, J., et al. (2010). Enzymatic AND logic gates operated under conditions characteristic of biomedical applications. *J. Phys. Chem. B* **114**, 12166-12174.

[43] Privman, V., Strack, G., Solenov, D., Pita, M., Katz, E. (2008). Optimization of enzymatic biochemical logic for noise reduction and scalability: How many biocomputing gates can be interconnected in a circuit? *J. Phys. Chem. B* **112**, 11777-11784.

[44] Privman, V., Arugula, M.A., Halámek, J., Pita, M., Katz, E. (2009). Network analysis of biochemical logic for noise reduction and stability: A system of three coupled enzymatic AND gates. *J. Phys. Chem. B* **113**, 5301-5310.

[45] Privman, V., Halámek, J., Arugula, M. A., Melnikov, D., Bocharova, V., Katz, E. (2010). Biochemical filter with sigmoidal response: Increasing the complexity of biomolecular logic. *J. Phys. Chem. B* **114**, 14103-14109.

[46] Privman, V., Pedrosa, V., Melnikov, D., Pita, M., Simonian, A., Katz, E. (2009). Enzymatic AND-gate based on electrode-immobilized glucose-6-phosphate dehydrogenase: Towards digital biosensors and biochemical logic systems with low noise. *Biosens. Bioelect.* **25**, 695-701.

[47] Researchers from Clarkson University report details of new studies and findings in the area of biosensors and bioelectronics. *Electronics Newsweekly* (Atlanta, GA), Issue: Jan. 13, 2010, Page: 161,

(online version at http://www.verticalnews.com/article.php?articleID=3010969).

[48] Buchler, N.E., Gerland, U., Hwa, T. (2005). Nonlinear protein degradation and the function of genetic circuits. *Proc. Natl. Acad. Sci. USA* **102**, 9559-9564.

[49] Setty, Y., Mayo, A.E., Surette, M.G., Alon, U. (2003). Detailed map of a cis-regulatory

input function. *Proc. Natl. Acad. Sci. USA* **100**, 7702-7707.

[50] Alon, U. (2007). *An Introduction to Systems Biology. Design Principles of Biological Circuits*, Boca Raton, Florida: Chapman & Hall/CRC Press.

[51] Halámek, J., Zhou, J., Halámková, L., Bocharova, V., Privman, V., Wang, J., Katz, E. (2011). Biomolecular Filters for Improved Separation of Output Signals in Enzyme Logic Systems Applied to Biomedical Analysis. *Anal. Chem.* **83**, 8383-8386.

[52] Privman, V. (2011). Control of Noise in Chemical and Biochemical Information Processing. *Isr. J. Chem.* **51**, 118-131.

[53] Pita, M., Privman, V., Arugula, M.A., Melnikov, D., Bocharova, V., Katz, E. (2011). Towards Biochemical Filter with Sigmoidal Response to pH Changes: Buffered Biocatalytic Signal Transduction. *Phys. Chem. Chem. Phys.* **13**, 4507-4513.

[54] Rafael, S.P., Vallée-Bélisle, A., Fabregas, E., Plaxco, K., Palleschi, G., Ricci, F. (2012). Employing the Metabolic "Branch Point Effect" to Generate an All-or-None, Digital-like Response in Enzymatic Outputs and Enzyme-Based Sensors. *Anal. Chem.* **84**, 1076-1082.

[55] Moser, I., Jobst, G., Svasek, P., Varahram, M., Urban, G. (1997). Rapid liver enzyme assay with miniaturized liquid handling system comprising thin film biosensor array. *Sens. Actuators B* **44**. 377-380.

[56] Kato, G.J., McGowan, V., Machado, R.F., Little, J.A., Taylor, VI, J., Morris, C.R., Nichols, J.S., Wang, X., Poljakovic, M., Morris, Jr., S.M., Gladwin, M.T. (2006). Lactate dehydrogenase as a biomarker of hemolysis-associated nitric oxide resistance, priapism, leg ulceration, pulmonary hypertension, and death in patients with sickle cell disease. *Blood* **107**, 2279-2285.

[57] Halámek, J., Zavalov, O., Halámková, L., Korkmaz, S., Privman, V., Katz, E. (2012). Enzyme-Based Logic Analysis of Biomarkers at Physiological Concentrations: AND Gate with Double-Sigmoid "Filter" Response. *The Journal of Physical Chemistry B* **116**, 4457-4464.

[58] Zavalov, O., Bocharova, V., Privman, V., Katz, E. (2012). Enzyme-Based Logic: OR Gate with Double-Sigmoid Filter Response. *The Journal of Physical Chemistry B* **116**, 9683–9689.

[59] Kotoh, K., Enjoji, M., Kato, M., Kohjima, M., Nakamuta, M., Takayanagi, R. (2008). A new parameter using serum lactate dehydrogenase and alanine aminotransferase level is

useful for predicting the prognosis of patients at an early stage of acute liver injury: A retrospective study. *Compar. Hepatol.* **7**, 6-14.

[60] Khalili, H., Dayyeh, B.A., Friedman, L.S. (2010). In: *Clinical Gastroenterology: Chronic Liver Failure*, Ginès, P., Kamath, P.S., Arroyo, V. (Eds.) Humana Press, New York, 47-76.

[61] Zhou, J., Halámek, J., Bocharova, V., Wang, J., Katz, E. (2011). *Talanta* **83**, 955-959.

[62] Privman, M., Tam, T.K., Bocharova, V., Halámek, J., Wang, J., Katz, E. (2011). Responsive interface switchable by logically processed physiological signals: toward "smart" actuators for signal amplification and drug delivery. *ACS Appl. Mater. Interfaces* **3**, 1620-1623.

[63] Tokarev, I., Gopishetty, V., Zhou, J., Pita, M., Motornov, M., Katz, E., Minko, S. (2009). Stimuli-responsive hydrogel membranes coupled with biocatalytic processes. *ACS Appl. Mater. Interfaces* **1**, 532-536.

[64] Motornov, M., Zhou, J., Pita, M., Tokarev, I., Gopishetty, V., Katz, E., Minko, S. (2009). Integrated multifunctional nanosystem from command nanoparticles and enzymes. *Small* **5**, 817-820.

[65] Motornov, M., Zhou, J., Pita, M., Gopishetty, V., Tokarev, I., Katz, E., Minko, S. (2008). "Chemical transformers" from nanoparticle ensembles operated with logic. *Nano Lett.* **8**, 2993-2997.

[66] Bychkova, V., Shvarev, A., Zhou, J., Pita, M., Katz, E. (2010). Enzyme logic gate associated with a single responsive microparticle: Scaling biocomputing to microsize systems. *Chem. Commun.* **46**, 94-96.

[67] Bocharova, V., Tam, T.K., Halámek, J., Pita, M., Katz, E. (2010). Reversible gating controlled by enzymes at nanostructured interface. *Chem. Commun.* **46**, 2088-2090.

[68] Adler, A.J., Kristiakowsky, G.B. (1962). *J. Am. Chem. Soc.* **84**, 695-703.

[69] Bergmeyer, H.U. (1974). *Methods of Enzymatic Analysis*, 2$^{nd}$ ed., (Ed.) Academic Press, New York, Vol. **4**, 2066-2072.

[70] Craig, N.C., Kistiakowsky, G.B. (1958). Kinetics of Ester Hydrolysis by Horse Liver Esterase. *J. Am. Chem. Soc.* **80**, 1574-1579.

[71] Kyger, E.M., Riley, D.J., Spilburg, C.A., Lange, L.G. (1990). Pancreatic cholesterol esterases. 3. Kinetic characterization of cholesterol ester resynthesis by the pancreatic

cholesterol esterases. *Biochemistry* **29**, 3853-3858.

[72] Millero, F.J. (2009). Use of the Pitzer Equations to Examine the Dissociation of TRIS in NaCl Solutions. *J. Chem. Eng. Data*. **54**, 342-344.

[73] Koudelka, G.B., Hansen, F.B., Ettinger, M.J. (1985). Solvent isotope effects and the pH dependence of laccase activity under steady-state conditions. *J. Biol. Chem.* **260**, 15561-15565.

**Nomenclature**

| | | |
|---|---|---|
| ALT | | Alanine transaminase |
| LDH | | Lactate dehydrogenase |
| [E] | 0.5 U = 4.85×10$^{-5}$ mM | Esterase concentration |
| [A] | | Ethyl butyrate concentration |
| [B] | | Methyl butyrate concentration |
| [U] | | Butyric acid concentration |
| [T] | | Tris buffer concentration |
| TH$^+$ | | Protonated form of Tris buffer |
| [L] | 0.5 U = 3.78×10$^{-4}$ mM | Laccase concentration |
| [F] | 1 mM | K$_4$Fe(CN)$_6$ concentration |
| [P] | | K$_3$Fe(CN)$_6$ concentration |
| $x$ | | Logic input |
| $y$ | | Logic input |
| $z(x,y)$ | | Logic gate response surface |
| G6PDH | | Glucose-6-phosphate dehydrogenase |
| Glc6P | | D-glucose-6-phosphate |
| NAD$^+$ | | β-Nicotinamide adenine dinucleotide cofactor |
| NADH | | β-Nicotinamide adenine dinucleotide reduced cofactor |
| 6-PGluc | | 6-phospho-gluconic acid |
| α-KTG | 10 mM | α-Ketoglutaric acid |
| Ala | 200 mM | Alanine |
| Glu | | Glutamate |
| Pyr | | Pyruvate |
| Lac | | Lactate |

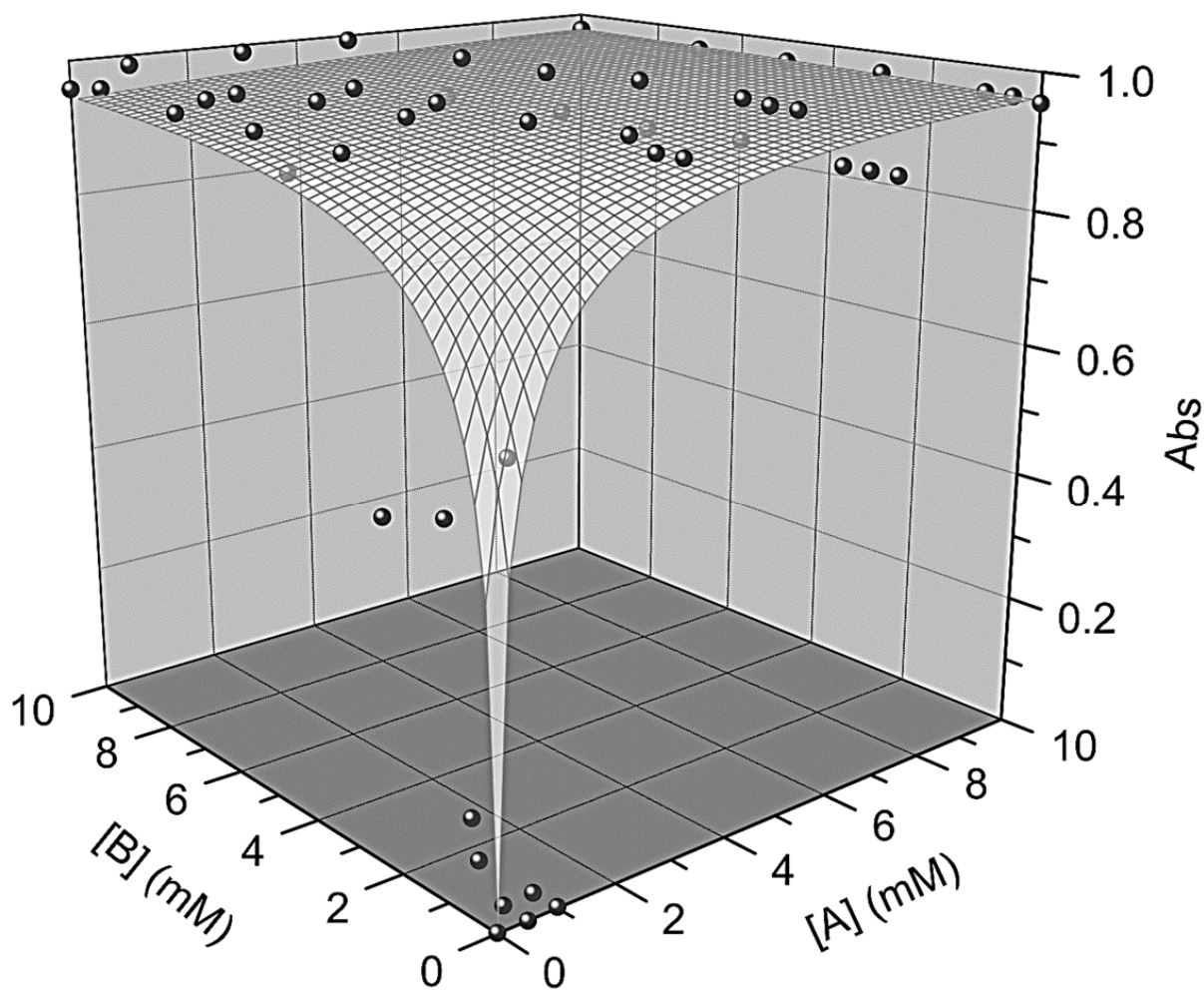

**Figure 1:** Realization of an **OR** gate for the biochemical system without filtering, by biochemical processes detailed in the text. Spherical symbols show the 7×7 grid of 49 experimental data points. These were measured at the gate time 800 s, for various values of the two input concentrations (ethyl butyrate [A] and methyl butyrate [B]), the logic point values for both of which were set at 0 and 10 mM concentrations (for **0** and **1**, respectively). The logic point values of the measured output, measured by the absorbance of the product (see text), are set by the gate function itself. The shown surface was calculated from the model described in the text.

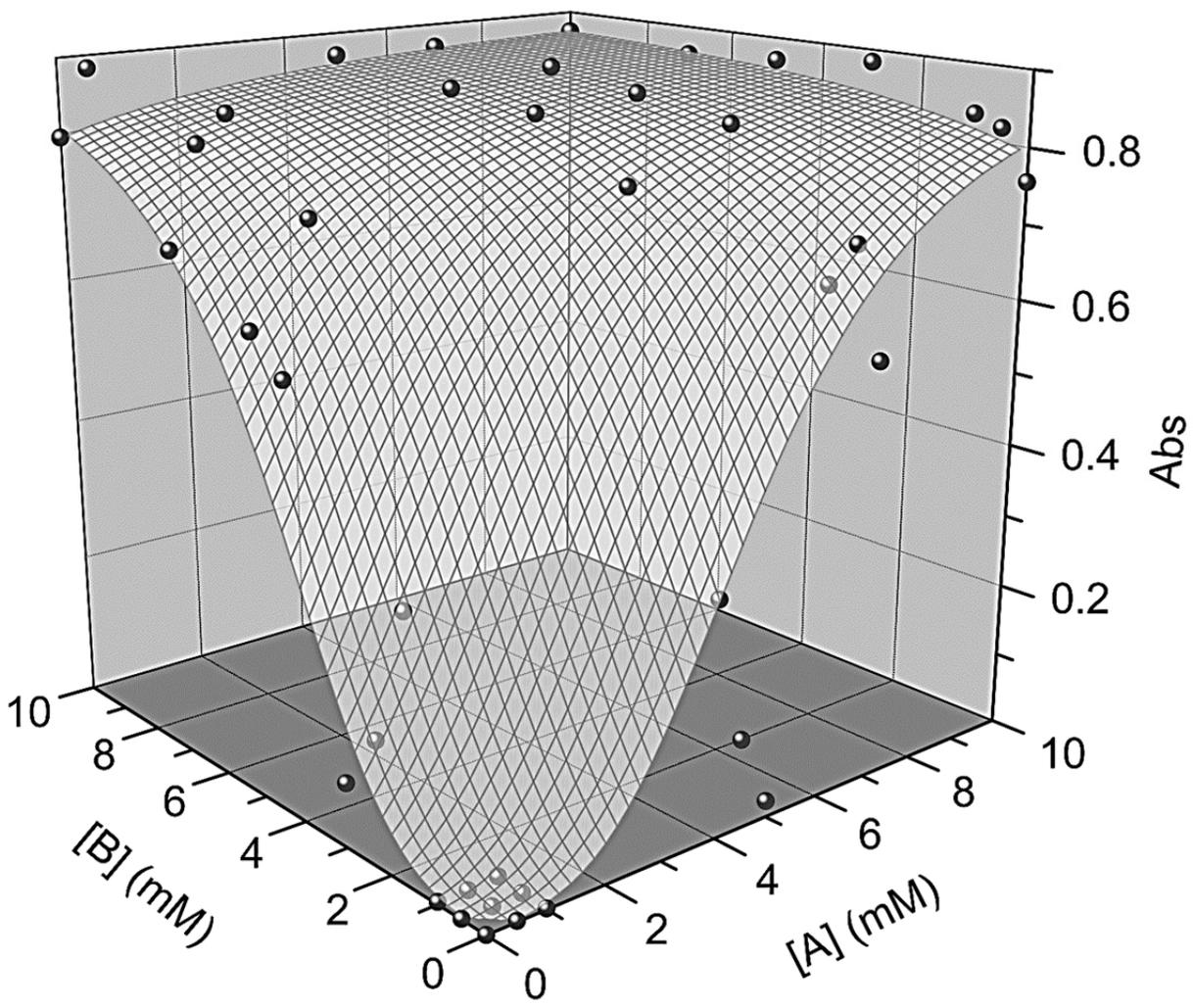

**Figure 2:** Experimental data and model-calculated surface for various values of the two input concentrations of ethyl butyrate and methyl butyrate (as in Figure 1) for the **OR** gate system with 4 mM buffer added initially (the optimal filtering process), gate time 800 s.

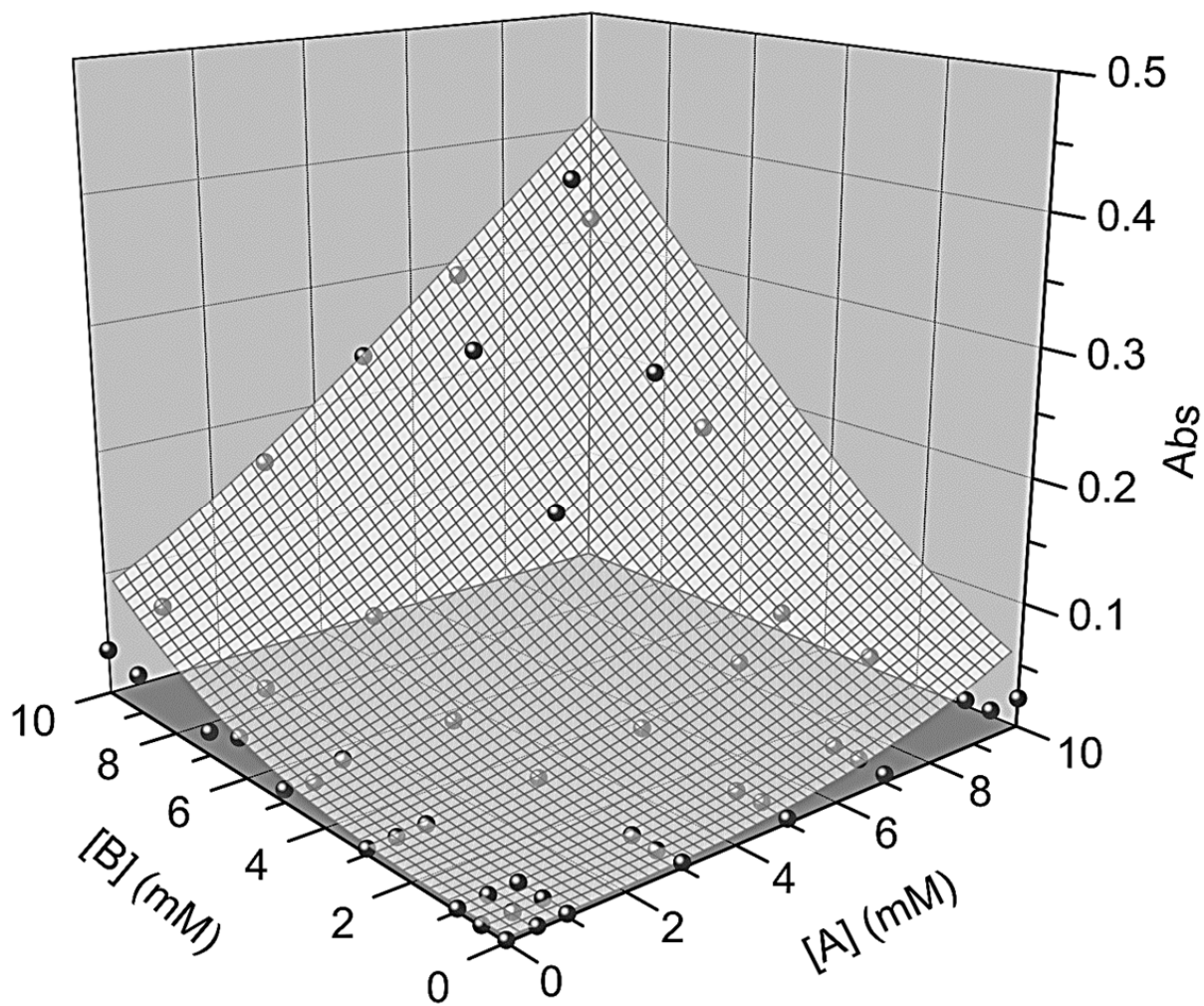

**Figure 3:** Experimental data and model-calculated surface (as in Figures 1 and 2) for the **OR** gate system with 8 mM buffer added initially, gate time 800 s. The **OR** function is not well realized in this over-filtered system.

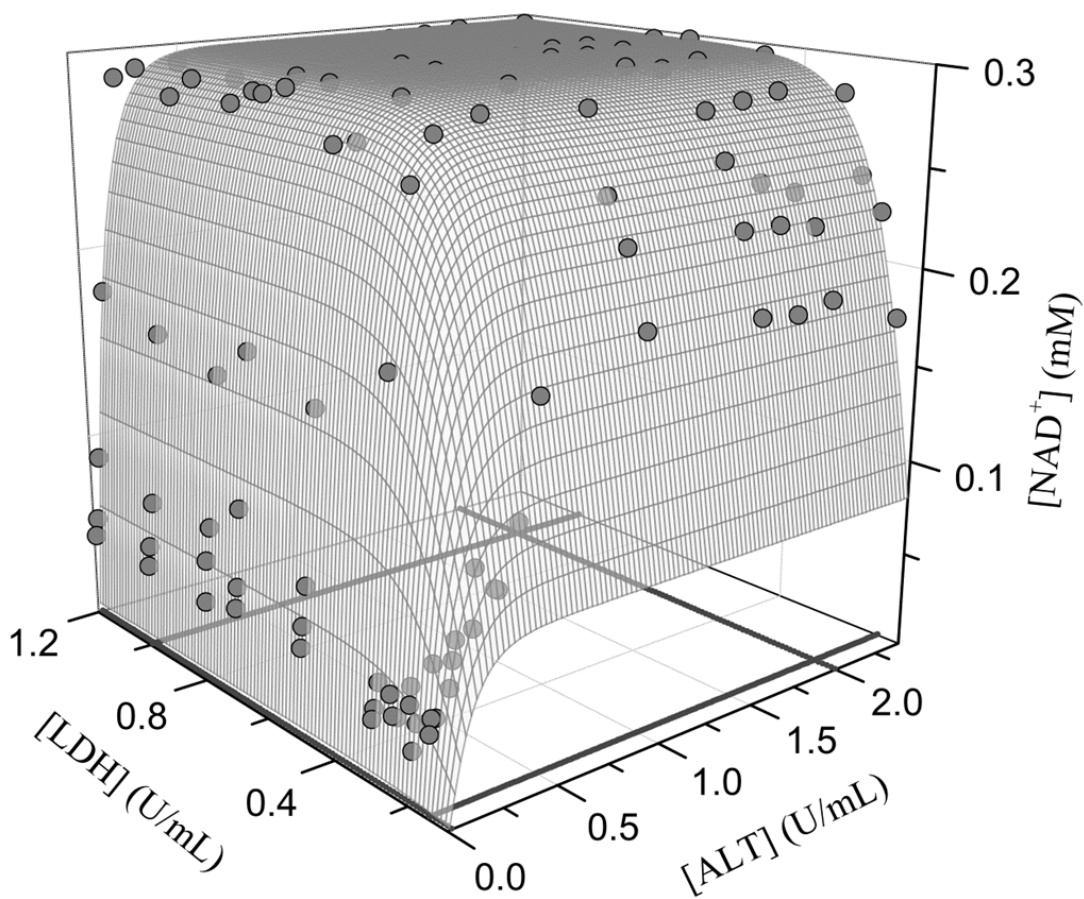

**Figure 4:** Realization of an **AND** gate for the biochemical system without the filter. Spherical symbols show the 108 (12×9) experimental data points. These were measured at the gate time $t = 600$ s, for various values of the two input enzyme concentrations as detailed in the text. The lines under the surface are drawn at the selected logic-**0** and **1** values of the inputs and thus delineate the "logic" range for mapping out the gate-response function. The surface was calculated from the model described in the text, with the parameters fitted based on the full time-dependent data set for the "filtered" system.

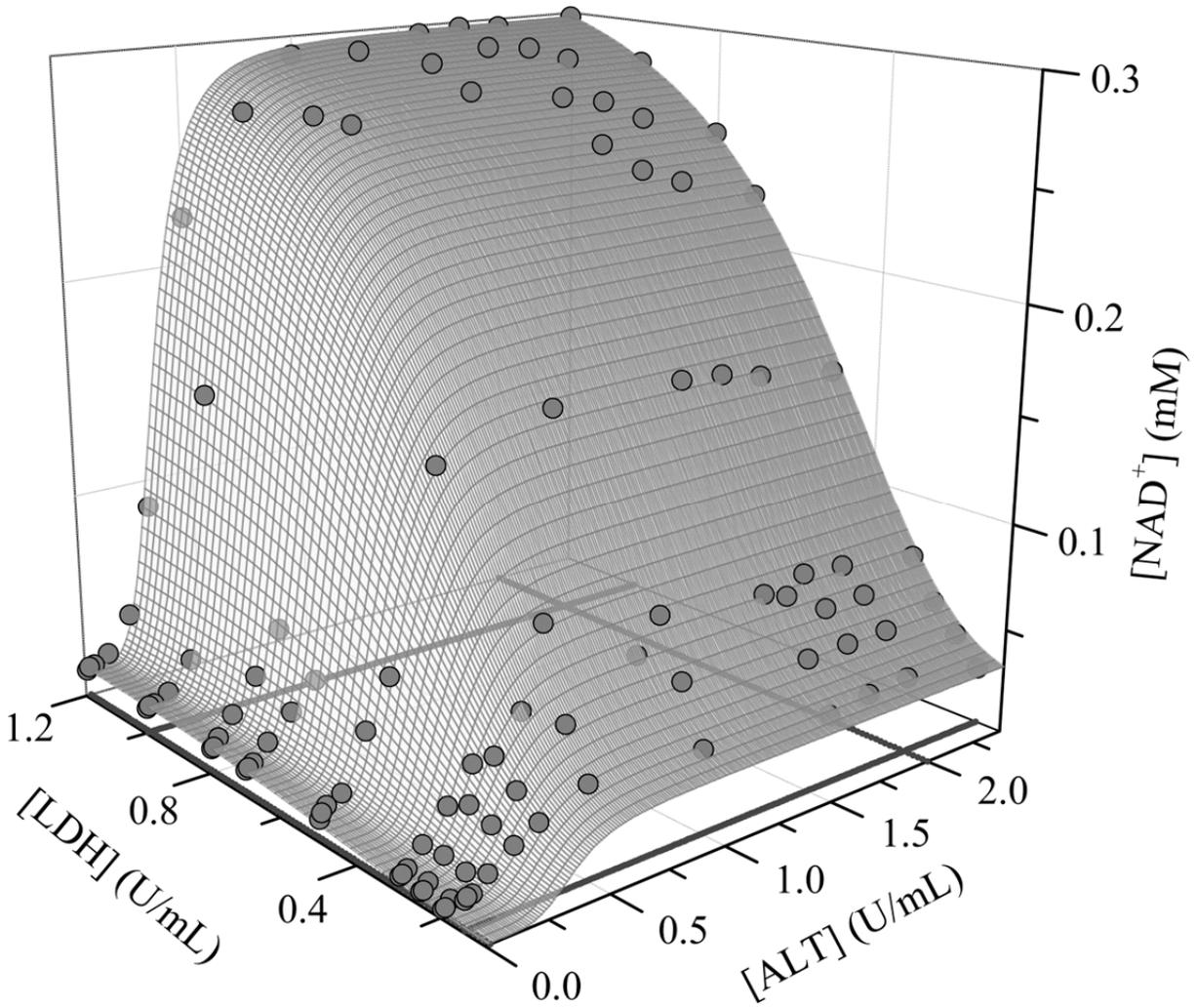

**Figure 5:** The **AND** function realization for the biochemical system with the filter process. Spherical symbols show the 108 (12×9) experimental data points, measured at the gate time $t = 600$ s. The lines under the surface mark the selected logic-**0** and **1** values of the inputs. The surface was calculated from the model described in the text, with the parameters fitted based on the full time-dependent data set for the "filtered" system.

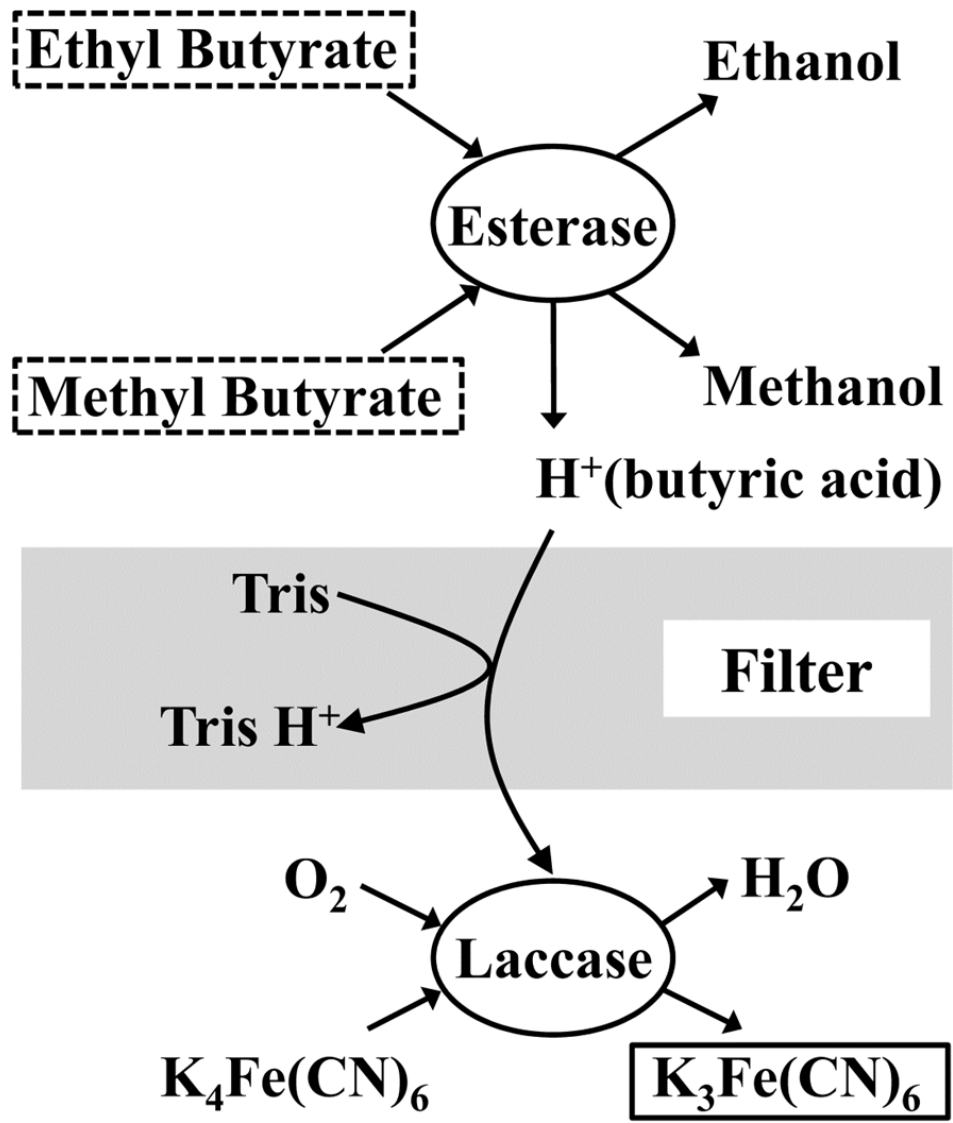

**Scheme 1.** Chemical and biochemical processes in the **OR** gate system. All the notations are defined in the text. The two inputs are marked by the dashed-line boxes, whereas the output is marked by the solid box.

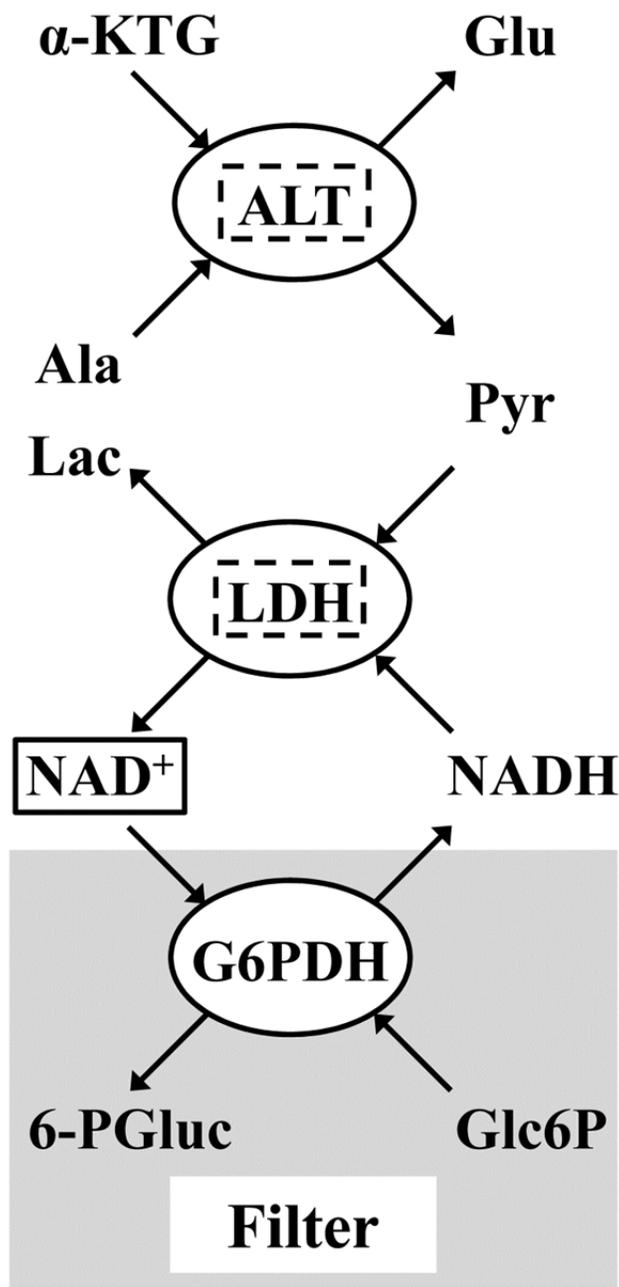

**Scheme 2.** Biocatalytic cascade realizing the **AND** gate function with the filter process added. All the notations are defined in the text. The two inputs are marked by the dashed-line boxes, whereas the output is marked by the solid box.